\documentclass[12pt,a4paper]{article}

\usepackage{amsmath}
\usepackage{amssymb}
\usepackage{mathrsfs}
\usepackage{graphicx}
\usepackage{enumerate}

\def\C{\mathbb{C}}
\def\Z{\mathbb{Z}}
\def\R{\mathbb{R}}
\def\P{\mathbb{P}}
\def\H{\mathbb{H}}
\def\RP{\mathbb{R}\mathbb{P}}

\usepackage[margin=1in]{geometry}

\usepackage[colorlinks=true,linkcolor=blue,citecolor=blue,urlcolor=blue]{hyperref}

\begin{document}

\title{Holography on Non-Orientable Surfaces}
\author{Alexander Maloney$^{1}$\footnote{maloney@physics.mcgill.ca}{ } and Simon F. Ross$^{2}$\footnote{s.f.ross@durham.ac.uk} \\  \bigskip \\
$^{1}$McGill Physics Department, 3600 rue University, \\ Montr\'eal, QC H3A 2T8, Canada
\\
\\
$^{2}$Centre for Particle Theory, Department of Mathematical Sciences \\ Durham University\\ South Road, Durham DH1 3LE}

\maketitle

\vspace{-1em}

\abstract{
We consider the holographic computation of two dimensional conformal field theory partition functions on non-orientable surfaces. We classify the three dimensional geometries that give bulk saddle point contributions to the partition function, and find that there are fewer saddles than in the orientable case.  For example, for the Klein bottle there is a single smooth saddle and a single additional saddle with an orbifold singularity.  We argue that one must generally include singular bulk saddle points in order to reproduce the CFT results.  We also discuss loop corrections to these partition functions for the Klein bottle.
}

\clearpage

\tableofcontents

\section{Introduction}

The AdS/CFT correspondence states that the quantum gravity with asymptotically AdS boundary conditions is equivalent to a conformal field theory living on the boundary of AdS \cite{Maldacena:1997re}.  This implies, for example, that the partition function of a bulk theory of gravity equals that of a conformal field theory in one less dimension.  This remarkable proposal is usually difficult to test directly.   An important exception, however, occurs in the case of three dimensional gravity, where a variety of exact computations are possible.  

We will study the partition function of three dimensional theories of gravity in Euclidean signature.  This partition function should, at least in principle, be regarded as an integral over a space of Euclidean bulk solutions with specified boundary conditions. 
The partition function will in general be a function of the sources for any bulk fields which are turned on at the asymptotic boundary.  In this paper we will focus on the metric degrees of freedom, and set other sources to be zero.  The partition function is then a function of the topology and the conformal structure of the asymptotic boundary.  Our goal is to compute these functions, and to thus use CFT consistency conditions to constrain possible bulk theories of  gravity.

These computations are typically carried out in the case where the boundary is a Riemann surface.  The simplest case is when the boundary is the sphere $S^2$.  In this case there is a single semi-classical saddle that contributes to the sum over geometries: Euclidean AdS$_3$, i.e. hyperbolic space.  Unless sources for bulk fields are turned on, the result has a trivial interpretation in the boundary CFT; the CFT partition on $S^2$ is just a number, which can be interpreted as the norm of the ground state of the theory.
A more interesting case is when the boundary is the torus $T^2$.  The torus partition function encodes the spectrum of the dual CFT, which is then highly constrained by the modular invariance on the torus.  This was used to argue that the bulk gravity theory must include an infinite number of classical saddle-points \cite{Maldacena:1998bw, Dijkgraaf:2000fq,Manschot:2007ha}.  Higher genus Riemann surfaces can also be considered, although the constraints of higher genus modular invariance are more difficult to extract (see e.g.\cite{Yin:2007gv}).

We will consider instead the partition function on non-orientable surfaces, focusing on the case of $\R\P^2$ and the Klein bottle $K^2$.  The structure of conformal field theory on non-orientable surfaces is well studied, and appears in the study of string worldsheet theories in the presence of orientifolds (see e.g.
\cite{Fioravanti:1993hf,Pradisi:1995pp} and \cite{Stanev:2001na}, \cite{Blumenhagen:2009zz} for reviews).

We will describe and classify the bulk saddle point geometries which contribute to these partition functions.  This discussion will closely follow the construction in the orientable case. 
We will find, however, far fewer smooth bulk geometries than in the orientable case.   For $\R\P^2$ there are no smooth bulk geometries.  For $K^2$, there is only one smooth saddle; this should be compared to the infinite number of smooth saddles in the $T^2$ case.  The classical action of this saddle point contributes to the partition function of AdS$_3$ gravity with $K^2$ boundary conditions.  We will also discuss the perturbative corrections to saddle point geometries, comparing the bulk one-loop determinant to CFT expectations.

We find that, in the regime of large central charge where the semi-classical calculation can be trusted, including just the smooth saddle-points does not reproduce the expected CFT behaviour.  We will find a class of non-smooth saddles that obey the desired boundary conditions, which contain simple $\Z_2$ orbifold singularities, and argue that these singular saddles must be included in the gravity path integral if we are to reproduce typical CFT behaviour. We also consider the one-loop determinant about both the singular and smooth saddles, and find that they reproduce general CFT expectations. 
\footnote{It is important to note that many theories can only be defined on orientable surfaces, so the non-orientable partition functions vanish. A notable example is chiral gravity \cite{Li:2008dq}, for which the non-orientable partition function appears to vanish identically in both the boundary and the bulk.  So our conclusions do not, for example, contradict the proposal of \cite{Maloney:2009ck} that chiral gravity can be regarded as a sum over only smooth geometries.  In this paper we consider only cases -- such as pure Einstein gravity without fermions -- where the non-orientable contributions are not manifestly zero.}

These observations are important for our understanding of the holographic dictionary for general two-dimensional CFTs, and also for attempts to construct theories of pure gravity. 
If one wishes to define a theory of pure gravity in AdS$_3$, whose degrees of freedom include only the metric, the most natural definition is that the path integral should include only smooth geometries.  This is because the non-smooth geometries which locally solve the equations of motion typically have orbifold singularities which are associated with new degrees of freedom. This proposal was advocated in
\cite{Witten:2007kt,Maloney:2007ud, Keller:2014xba}, who defined pure gravity as the theory whose Euclidean path includes only smooth saddles.  These authors computed, for the case of the torus, the partition function of general relativity exactly including these smooth geometries.  For large central charge, which would describe the dual of semi-classical gravity, the result did not satisfy the axioms of a conformal field theory; the resulting partition function could not be interpreted as a finite temperature partition function of a theory with a positive definite spectrum.  

The computation of the torus partition function outside of the semi-classical regime was considered in \cite{Castro:2011zq}.  The torus partition function, computed as a sum over smooth geometries, was found to match that of a minimal model CFT for certain ${\cal O}(1)$ values of the central charge.   This was interpreted as evidence that pure gravity might exist as a proper quantum mechanical theory for certain highly quantum mechanical values of the cosmological constant.  For example, the Ising model CFT with $c=1/2$ was conjectured to be dual to general relativity in a highly quantum regime, where the curvature of AdS is Planck scale.  We are now in a position to test this conjecture for the non-orientable saddles $\R\P^2$ and $K^2$. Our conclusion is that the paucity of smooth bulk saddle points implies that the original conjecture -- that these minimal models are obtained by a sum only over smooth saddle points -- must be modified.

In the next section, we discuss the construction of the bulk saddle-points contributing to the partition function for a non-orientable surface in general, and describe the resulting saddles in detail for $\RP^2$ and the Klein bottle $K^2$. In section \ref{cft} we discuss the behaviour of CFT partition functions on the Klein bottle in certain limits of the Klein bottle modulus, and argue that these indicate that we must typically include as saddle points geometries which contain orbifold singularities. In section \ref{oneloop}, we carry out the one-loop determinant calculation from the bulk point of view, and show that this matches field theory expectations.   We will use both a method of images calculation and a mode sum, and see that these agree with the CFT expectations for both the singular and smooth saddles. There is an interesting subtlety in the calculation for the smooth case.

\section{Classical Bulk Saddles}
\label{classical} 

We will consider the partition function of three dimensional gravity in Euclidean signature with asymptotically AdS boundary conditions.  This partition function will depend on the topology and conformal structure of the geometry at the asymptotic boundary, which we denote $\Sigma$. We will describe  the solutions of three dimensional Einstein gravity where the boundary $\Sigma$ is non-orientable.  These geometries give saddle point contributions to the partition function.

We will focus on the case of pure Einstein gravity, when there are no additional bulk fields present.  In this case the solutions are quotients of Euclidean AdS$_3$ (i.e. $\H^3$), so are easy to classify. In pure Einstein gravity these  are the only solutions.  Even if other bulk fields are present, these quotients will still be solutions, since they locally solve the same equation of motion as empty AdS. However, in the more general case additional solutions may be present.

We will begin by describing the general features of the solutions, before moving on to specific examples.

\subsection{Filling in Non-Orientable Surfaces}

Topologically, every non-orientable surface $\Sigma_g$ can be written as the connected sum of $g$ copies of real projective space 
\begin{equation}
\Sigma_g = \RP^2 \# \dots \#\RP^2
\end{equation}
where the integer $g\ge 1$ is the genus of the surface.  The process of taking a connected sum with $\RP^2$ (removing a disk and gluing in a cross-cap) has a natural CFT interpretation in terms of the cross cap state, which will be useful in the next section.  

We wish to find locally hyperbolic three-manifolds whose conformal boundary is $\Sigma_g$.  Any such bulk geometry must be a quotient of the form $\H^3/\Gamma$, where $\Gamma$ is a discrete subgroup of the $SL(2,\C)$ isometry group of $\H^3$.

In constructing the saddles, we will find it convenient to use the fact that any non-orientable manifold can be represented as the $\Z_2$ quotient of an orientable manifold: the
surface $\Sigma_g$ can be represented as the $\mathbb Z_2$ quotient of an orientable surface $\hat \Sigma_g$.  
Here $\hat \Sigma_g$ is the natural double cover (called the orientable double cover) where we simply take two copies of each point on $\Sigma$, one for each orientation. ${\hat \Sigma}_g$ is orientable by construction, and has a $\mathbb Z_2$ symmetry interchanging the two orientations. 
The statement that $\Sigma_g = {\hat \Sigma}_g/\Z_2$ is the natural generalization of the observation that $\RP^2$ is the quotient of the sphere by the antipodal map, $S^2/\Z_2$.

This observation applies in the bulk as well, and allows us to related the bulk saddles with non-orientable boundary to the more familiar case of bulk saddles with orientable boundary.  We first note that any 3-manifold with non-orientable boundary must be non-orientable itself.  Thus any any bulk geometry with $\Sigma_g$ boundary is the $\Z_2$ quotient of an orientable manifold whose boundary is the orientable double cover $\hat \Sigma_g$.  There is no guarantee, however, that the bulk saddles obtained in this way will be smooth.

This fact also allows us to describe the moduli space of conformal structures of a general non-orientable surface in terms of the more familiar moduli space of Riemann surfaces.
The moduli space of $\Sigma_g$ is just the subspace of the moduli space of $\hat \Sigma_g$ which preserves the $\mathbb Z_2$ symmetry.\footnote{For example, the Klein bottle has a single real modulus, and it is obtained as the quotient of a rectangular torus $t \sim t + \beta, \phi \sim \phi + 2\pi$ for real $\beta$ by the $\mathbb Z_2$ symmetry $t \to -t, \phi \to \phi + \pi$. More general tori do not arise as the double cover of a non-orientable surface.} The bulk saddle point action will be a function of this moduli space.

\subsection{$\R\P^2$}

Real projective space, $\R \P^2$ is the quotient of the sphere by the antipodal map: $\R\P^2 = S^2 / \Z_2$. 
The bulk saddles 
are of the form $\H^3/\Z_2$.
In order to specify the bulk saddle point uniquely, we must therefore choose a $g\in SL(2,\C)$ with $g^2=1$.  Every such $g$ has a single fixed point in the bulk, and different choices of $g$ correspond to different choices of fixed point in the bulk.  These saddles can be visualized by starting with the $\H^3$ which has an $S^2$ boundary, and extending the antipodal map on $S^2$ into the interior of $\H^3$.  Of course, one can always conjugate this antipodal map by an isometry on $\H^3$ (i.e. a conformal transformation on $S^2$), which has the effect of moving the bulk fixed point and changing the resulting element $g$.
 
We note in particular that since every $g\in SL(2,\C)$ has a fixed point there are no smooth saddles for $\RP^2$.  So if the bulk path integral includes a sum only over smooth saddles there will be no contribution to the $\RP^2$ partition function $Z_{\RP^2}$.
This already is a bit mysterious from the CFT point of view, since typically $Z_{\RP^2}$ is non-zero.  

\subsection{Klein Bottle}
\label{kleins}

\begin{figure}
\centering 
\includegraphics[width=9cm]{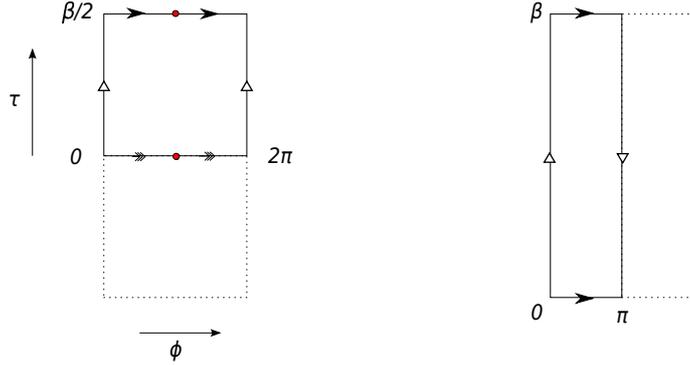}
\caption{The Klein bottle can be thought of as the quotient of a rectangular torus by the $\mathbb Z_2$ action $(t, \phi) \sim (-t, \phi+\pi)$. There are two natural fundamental regions for this identifications. On the left, we have a representation as the propagation between two cross-caps. On the right, we have the alternative representation with an orientation-reversing identification of the two sides.} \label{kb}
\end{figure}

The next example is the Klein bottle $K^2 = \RP^2 \# \RP^2$.  This geometry is the  $\Z_2$ quotient of the rectangular torus: $K^2 = T^2/\Z_2$.  Explicitly, we quotient the rectangular torus $t \sim t + \beta, \phi \sim \phi + 2\pi$ by the $\mathbb Z_2$ symmetry $t \to -t, \phi \to \phi + \pi$, as depicted in figure \ref{kb}. 
We can identify saddle points with Klein bottle boundary conditions in the bulk by looking for saddle points for the torus which are invariant under this $\mathbb Z_2$ action. 

We will begin by describing the bulk saddle points with torus boundary.  These were classified by \cite{Maloney:2007ud}, who showed that the only smooth saddles are handlebodies which are topologically equivalent to a solid donut.  
 These handlebodies are labelled by which non-contractible cycle in the torus becomes contractible in the bulk. Explicitly, the torus has  $H^1(T^2)=\mathbb Z \oplus \mathbb Z$, with a natural basis consisting of an $a$ cycle along $t$ and a $b$ cycle along $\phi$. Any combination $m a + n b$ can become contractible in the bulk, so there is an infinite family of  saddles in the bulk, labelled by two integers $m,n$. 
 Since the $\mathbb Z_2$ is a symmetry of the boundary, a given bulk saddle will either be invariant under it or will be exchanged with some other saddle.
 Under the orientation-reversing $\mathbb Z_2$ action, the $a$ cycle is odd and the $b$ cycle is even. Thus, a cycle $m a + n b$ is exchanged with the cycle $-m a + n b$ under the $\mathbb Z_2$, and the corresponding bulk handlebodies are also exchanged.  We conclude that there are only two bulk handlebodies which are invariant under the $\mathbb Z_2$: the one where the $a$ cycle is contractible and the one where the $b$ cycle is contractible. 

This allows us to construct the Klein bottle saddle points as quotients of these two geometries.  The Klein bottle has $H^1(K^2)=\mathbb Z \oplus \mathbb Z_2$, where the $\mathbb Z$ is the $a$ cycle, along the $t$ direction, and the $\mathbb Z_2$ is the $b$ cycle, along the $\phi$ direction.  When the $a$ cycle is contractible in the bulk, the $\mathbb Z_2$ symmetry acts without fixed points, and we obtain a smooth quotient. For the torus this saddle is the non-rotating BTZ black hole, and for the Klein bottle it is the Euclidean version of the $\R\P^2$ geon studied in \cite{Louko:1998hc}. When the $b$ cycle is contractible in the bulk, the saddle for the torus is thermal AdS. This has fixed points for the $\mathbb Z_2$ symmetry $t \to -t, \phi \to \phi + \pi$ at the origin in the spatial slices at $t=0, \beta/2$. So the resulting saddle for the Klein bottle has two $\mathbb Z_2$ fixed points. 

For later use, we now discuss these two solutions explicitly. We work in the upper-half plane picture of Euclidean AdS$_3$, that is the hyperbolic three-space $\mathbb H^3$, where
\begin{equation}
ds^2 = \frac{dy^2 + dz d\bar z}{y^2}, 
\end{equation}
with $y\in (0, \infty)$, and $z$ is a complex coordinate on the planes of constant $y$. We will also find it convenient to introduce the coordinates $\rho, \theta, \phi$ where 
\begin{equation}
y = e^\rho \sin \theta, \quad z = e^\rho \cos \theta e^{i \phi},  
\end{equation}
with $\rho \in (-\infty, \infty)$, $\theta \in (0, \pi/2)$ and $\phi \in (0, 2\pi)$. The orbifold by $\gamma: (y, z) \to (e^\beta y, e^\beta z)$ makes the boundary at $z=0$ into a rectangular torus. In the $\rho, \theta, \phi$ coordinates, this acts as $\rho \to \rho + \beta$, so a fundamental region is $\rho \in (0, \beta)$. If we write $z = e^{t + i \phi}$, where $t$ is the Euclidean time coordinate on the boundary, this is the torus with $t \sim t + \beta$, $\phi \sim \phi + 2\pi$, and the bulk solution is thermal AdS$_3$, as the $\phi$ circle becomes contractible in the bulk. If we write $z = e^{\frac{\beta}{2\pi}(\phi + i t)}$, it is the torus with $t \sim t + 4\pi^2/\beta$, $\phi \sim \phi + 2\pi$, and the bulk solution is the  non-rotating BTZ black hole, where the $t$ circle becomes contractible in the bulk. 

The saddles for the Klein bottle are obtained by taking the $\mathbb Z_2$ quotient $t \to - t$, $\phi \to \phi + \pi$. In the first case, where  $z = e^{t + i \phi}$, this corresponds to the action $z \to - \bar z^{-1}$. In the second case,  with $z = e^{\frac{\beta}{2\pi}(\phi + i t)}$, it is $ z \to e^{\beta/2} \bar z$. In the first case the modular parameter of the Klein bottle is $l = \beta/4\pi$, while in the second case it is $l = \pi/\beta$.

Each of these can be extended to a discrete (orientation-reversing) isometry of $\mathbb H^3$: the former is 
\begin{equation} \label{sigma}
\sigma: \quad z \to - \frac{z}{|z|^2 + y^2}, \quad y \to \frac{y}{|z|^2 + y^2} , 
\end{equation}
while the latter is 
\begin{equation} \label{kappa} 
\kappa: \quad z \to e^{\beta/2} \bar z, \quad y \to e^{\beta/2} y .
\end{equation}
In the former case $\sigma^2$ is the identity, so we obtain a space with a Klein bottle boundary by taking $\mathbb H^3/\Gamma$ where $\Gamma$ is the group generated by $\sigma$ and $\gamma$, 
$\Gamma = \{ \gamma^n, \sigma \gamma^n | n \in \mathbb Z \}$. Thus in this case $\Gamma \simeq \mathbb Z \times \mathbb Z_2$. A fundamental region for the identification $\mathbb H^3/\Gamma$ is $\rho \in (0, \beta/2)$. The quotient has fixed points at $z=0$ and $y = 1, e^{\beta/2}$; that is $\theta = \pi/2$, $\rho = 0, \beta/2$. 

In the latter case, $\kappa^2 = \gamma$, so we obtain a space with a Klein bottle boundary by taking $\mathbb H^3/\Gamma$ where $\Gamma$ is simply the group generated by $\kappa$. Thus $\Gamma = \{ \kappa^n | n \in \mathbb Z \}$. Thus in this case $\Gamma \simeq \mathbb Z $, and the quotient is freely acting in the bulk. A convenient fundamental region for the identification $\mathbb H^3/\Gamma$ is $\rho \in (0, \beta), \phi \in (0, \pi)$. This is simply the Euclidean geon geometry.   

Unlike in the torus case, where the infinite family of saddles gave the bulk partition function as an infinite sum, which could be interpreted as a sum over the modular group \cite{Dijkgraaf:2000fq}, the partition function for the Klein bottle has only two contributions. This in itself does not obviously lead to problems: the dominant contribution for all values of the modulus for the rectangular torus is either the thermal AdS or the non-rotating BTZ, which are precisely the two saddles which descend to saddles of the Klein bottle. 

However, if we restrict to smooth saddles, the Klein bottle partition function will be simply given by the unique contribution from the Euclidean geon (the quotient of non-rotating BTZ), so 
\begin{equation}
Z_K \approx Det(\Delta) e^{-I_g},  
\end{equation}
where $Det(\Delta)$ represents the one-loop determinant of bulk fields around this saddle-point, to be discussed later, and $I_g$ is the action of the Euclidean geon spacetime.  As a result, the partition function would not exhibit any analogue of the Hawking-Page transition as we vary the modulus.  We will see in the next section that this result is highly non-generic from the CFT point of view. We will argue subsequently that we recover the expected behaviour by including the saddle with fixed points. In section \ref{oneloop} we will probe this further by considering the one-loop contributions. 

\subsection{Other Examples}

Although we will focus on $\RP^2$ and $K^2$, we will also comment briefly on more complicated surfaces.  For the non-orientable surfaces $\Sigma_g=\RP^2 \#\dots\#\RP^2$ with $g\ge2$ the construction of the bulk saddle points is more complicated.  The orientable double cover $\hat \Sigma_g$ is now a Riemann surface with genus $\ge 2$.
For example, for $\Sigma_3=\RP^2\# \RP^2 \# \RP^2 = T^2 \# \RP^2$ the double cover is a genus two surface. 

As in the Klein bottle case, the saddles can be characterized by stating which cycles in $\Sigma_g$ become contractable in the bulk.  A natural basis of cycles for $\Sigma_g$ contains $a$ cycles $a_i$ and $b$ cycles $b_j$ with intersection numbers $a_i \cap b_j = \delta_{ij}$.  For the double cover $\hat \Sigma_g$ we can choose a basis of $a$ and $b$ cycles such that each cycle is either odd or even under the orientation reversing $\mathbb Z_2$.

The easiest way to construct bulk saddle points is by quotienting those for $\hat \Sigma_g$.  The simplest such solutions are handlebodies, which are labelled by a choice of half of the non-contractible boundary cycles which become contractible in the bulk. If these contractible cycles get exchanged with some other cycles under the  $\mathbb Z_2$ action (because they are a combination of odd and even cycles) then the corresponding handlebodies are similarly exchanged. The only handlebodies which are invariant under the $\mathbb Z_2$ action are those where the contractible cycles are each either odd or even under the $\mathbb Z_2$ action. It is possible to have some odd contractible cycles and some even ones, but not contractible cycles which are some combination of odd and even cycles. Thus, while the set of possible handlebody solutions for the non-orientable surface is much smaller than for the orientable double cover, for higher genus, where there are more than one odd or more than one even cycle, there will be infinite families of saddles for the non-orientable surface.  The situation is therefore much more complicated than the Klein bottle case. We conjecture that the bulk saddles which do not have the $\mathbb Z_2$ symmetry will all be subdominant in the moduli space of $\mathbb Z_2$-invariant orientable surfaces, as in the torus case, but we have not checked this explicitly. 

In addition to these handlebody geometries, there are also have non-handlebody saddles. As in the orientable cases, non-handlebody solutions for the non-orientable boundaries can be simply constructed by taking a two-boundary wormhole as in \cite{Maldacena:2004rf} and quotienting by a $\mathbb Z_2$ action which includes interchanging the two boundaries.  For example, let us consider the case where the surface $\Sigma_g$ has a freely acting $\Z_2$ symmetry, denoted $\sigma$.  This lifts to a symmetry of the orientable double cover $\hat \Sigma_g$. We can then consider the two-boundary solution $dy^2 + \cosh^2 y d{\hat \Sigma}^2_g$, which has topology ${\hat \Sigma}_g \times \R$. The quotient by $\sigma$ composed with $y \to -y$ is a non-handlebody saddle for $\hat \Sigma_g$: it has $y \in (0, \infty)$, with a copy of $\hat \Sigma_g$ at each $y >0$ and $\hat \Sigma_g/\sigma$ at $y=0$.  Taking the non-orientable quotient then leads to a saddle with $\Sigma_g$ boundary. 

If we include all the saddles, both the smooth ones and the ones with fixed point singularities, there is no particular reason to be interested in these non-handlebody solutions, as they are conjectured to be sub-dominant relative to the dominant handlebodies \cite{Yin:2007at}. If we were to want to consider just smooth solutions, the non-handlebodies could be important, as they might dominate over the handlebody we are allowed to keep. However, more recent work suggests the non-handlebodies are subdominant compared to all handlebody solutions \cite{higherZ}. Hence we will not consider the non-handlebodies further.

\section{CFT Interpretation \& the Crosscap State}
\label{cft}

We now discuss the CFT partition functions on closed non-orientable surfaces.  We will first review a few relevant facts about CFTs on non-orientable surfaces, which are typically studied in the context of orientifolds.  We then compare these CFT expectations to the bulk saddle-points described in the previous section. 

\subsection{CFT Partition Functions on $\RP^2$ and $K^2$}

Conformal field theory partition functions on Riemann surfaces are well studied.  Although they are typically difficult to compute exactly, they are in principle determined uniquely once the three-point function coefficients $C_{ijk}$ and the torus one-point functions are determined \cite{Moore:1989vd}.
Similarly, CFT partition functions on non-orientable surfaces are uniquely determined only once one specifies an additional piece of data: the 
one point functions  on $\RP^2$
\cite{Fioravanti:1993hf} (as reviewed in \cite{Blumenhagen:2009zz}). 
In particular, if we write $\RP^2$ as the $z$-plane quotiented by the involution $I(z) = - \bar z^{-1}$, whose fundamental region is the unit disc, the one-point functions of primary operators (with $h = \bar h$) are constrained by conformal invariance to be \cite{Fioravanti:1993hf}
\begin{equation} \label{ccone}
\langle \phi_k \rangle_c = \frac{\Gamma_k}{(1 + z \bar z)^{2h}}. 
\end{equation}
Here $k$ labels the primary operators in the theory, and the coefficients $\Gamma_k$ are the new data we must specify to define the CFT on non-orientable surfaces.

CFTs on non-orientable surfaces can be understood by regarding a non-orientable surface as constructed by gluing cross-caps into a closed orientable surface.  The CFT path integral on the cross-cap defines a state, $|C\rangle$, in the Hilbert space of the CFT on a circle.  The state $|C\rangle$ can be constructed explicitly in terms of the data $\Gamma_k$.
In particular, because the cross-cap is constructed from an antipodal identification on the disk (which takes $L_n \to (-1)^n {\bar L}_n$), it must satisfy
\begin{equation} \label{cccond}
(L_n - (-1)^n \bar L_{-n}) | C \rangle =0. 
\end{equation}
For rational CFTs, this can be solved explicitly in terms of Ishibashi states,  much like for boundary states in boundary conformal field theory. The cross-cap Ishibashi states are defined by summing over the descendents of a given primary, pairing the elements of the chiral and anti-chiral Verma modules, so 
\begin{equation}
| C_i \rangle \rangle =  \sum_{\vec m} (-1)^{\sum_j m_j} |\phi_i, \vec m \rangle \otimes U  |\bar \phi_i, \bar \vec m \rangle,
\end{equation}
where $\vec m$ denotes the state constructed by acting with the each raising operator $L_{-j}$ $m_j$ times on the primary, and $U$ is an anti-unitary operator (see \cite{Blumenhagen:2009zz} for details). These cross-cap Ishibashi states form a basis for the solutions of \eqref{cccond}, so the cross-cap state is a linear combination of them. 

Non-orientable partition functions can be interpreted in terms of insertions of this cross-cap state.
For example, $\RP^2$ is obtained by gluing a disk onto a cross-cap.
This means that equation (\ref{ccone})
can be interpreted 
as a transition amplitude between the vacuum state at the origin and the cross-cap state $| C \rangle$ on the boundary of the unit disc, $\langle \phi_k \rangle_c = \langle 0 | \phi_k | C \rangle$. In particular, the  $\RP^2$ partition function is simply the overlap between the vacuum state and the cross-cap, 
\begin{equation}
Z_{\mathbb RP^2}  = \langle 0 | C \rangle = \Gamma_{\mathbb I}. 
\end{equation}

The one-point functions \eqref{ccone} thus determine the cross-cap state 
\begin{equation}
|C \rangle = \sum_i \Gamma_i | C_i \rangle \rangle. 
\end{equation}
where the sum is over all primary states in the theory.
As in boundary CFT, these coefficients are restricted by a consistency condition \cite{Fioravanti:1993hf}
\begin{equation} \label{const}
\sum_k C_{ijk} \Gamma_k M \left[ \begin{array}{cc}  i  & \bar j \\ j & \bar i \end{array} \right]   = (-1)^{h_i - \bar h_i + h_j - \bar h_j} C_{ijp} \Gamma_p
\end{equation}
where $M$ is the usual conformal block transformation matrix.
This is a linear equation for the $\Gamma_k$; since there is one equation for each $p$, the system will determine $|C \rangle$ uniquely up to normalization. The cross-cap state is determined by the $\Gamma_k$, and hence the partition function on any non-orientable surface can be calculated given these $\Gamma_k$. 

Unlike the $\RP^2$ partition function, $Z_K(\beta)$ is a function of the modular parameter $\beta$ which contains detailed information about the cross cap state and the CFT spectrum. 
From the representation of the Klein bottle on the left panel of figure \ref{kb}, we see that the Klein bottle can be represented as the expectation value of the propagator $e^{-\beta H/2}$ between two cross-cap states.  
Explicitly, we have (see e.g. \cite{Blumenhagen:2009zz})
\begin{equation}
\label{zis}
Z_{K}(\beta)  = \langle C | e^{-\beta H/2} | C \rangle = \sum_i \Gamma_i^2 \langle \langle C_i | e^{-\beta H/2} | C_i \rangle \rangle = \sum_i \Gamma_i^2 \chi_i\left(i {\beta \over 2\pi}\right),  
\end{equation}
where $\chi_i$ is the character of the Virasoro representation built on the primary $\phi_i$.  
In the limit where $\beta$ is large this sum is dominated by the vacuum state if
$ \Gamma_{\mathbb I} \neq 0$, so that
\begin{equation} 
\label{largelpf}
Z_K(\beta\to\infty) \approx Z_{\mathbb RP^2}^2 \exp\left\{{\beta c\over 24}\right\}
\end{equation}
Here we have used the fact that $\Gamma_{\mathbb I}=Z_{\RP^2}$ and that the vacuum state has energy $H = -c/12$.

There is a second representation of the Klein bottle partition function, however, where we 
represent the Klein bottle as a the propagation over a dual channel with the insertion of a parity reversal operator, as in  
the right panel of figure \ref{kb}.  As usual, when doing so we must make the conformal transformation so that the $t$ circle has length $2\pi$, so the propagation will be over a distance $2\pi^2/\beta$. Thus we can write 
\begin{equation} \label{Ppf}
Z_{K}(\beta)  = \mathrm{Tr}(P e^{-2\pi^2 H/\beta }), 
\end{equation}
where $P$ is the parity operator, and the trace is over the Hilbert space of the CFT on the circle.  We note that parity-odd states contribute negatively to the trace.  
This  allows us to approximate $Z_K$ at small $\beta$, where the vacuum dominates in this channel:
\begin{equation}\label{smallbeta}
Z_K(\beta\to0) \approx \exp\left\{{\pi^2 c \over 6 \beta}\right\}
\end{equation}
We note that $Z_{\RP^2}$ does not appear in this expression; we have only assumed that the vacuum state is parity invariant.

\subsection{Comparison With Bulk Gravity}

Let us now compare these results with the bulk gravity expectations. In defining the partition function of three dimensional gravity as a sum over geometries, the first question is which geometries should be included. 

\subsubsection{Smooth saddles}

 The most natural supposition is that only smooth geometries contribute as saddle points.  We can now compare this with the CFT results described above.  We will restrict our attention to the semi-classical (large $c$) regime where the saddle point approximation is valid.

As we saw in section 2, there is no smooth bulk solution of Einstein gravity with $\RP^2$ boundary.  So if only smooth saddles are included, this would seem to set $Z_{\RP^2}=0$.  For the Klein bottle, we have one smooth saddle, the Euclidean continuation of the $\RP^2$ geon.  The classical action of this saddle is easy to compute: since the solution is the $\Z_2$ quotient of the BTZ black hole, the action just half the action of the non-rotating BTZ black hole.  This gives a contribution to the $K$ partition function
\begin{equation}
Z_{geon}(\beta) \approx  \exp\left\{{\pi^2 c \over 6 \beta}\right\}
\end{equation}
which correctly reproduces the small $\beta$ behaviour (\ref{smallbeta}) of the CFT.
  
The large $\beta$ behaviour of the CFT is however a more challenging test for the bulk gravity.  The action of the geon vanishes as $\beta\to \infty$, implying that at leading order in $c$ the partition function $Z_K$ should remain finite as $\beta\to\infty$, in contradiction with the exponential growth expected from (\ref{largelpf}).  However, we have already seen that including only smooth saddles led us to conclude that the  coefficient $Z_{\RP^2}=\Gamma_{\mathbb I}$ should vanish. The prediction for the large $\beta$ behaviour of $Z_K$ reinforces this requirement. From (\ref{zis}) we see that the bulk prediction that the action remains finite as $\beta\to \infty$ requires further that $\Gamma_i$ vanish for all operators with $dim(\phi_i)< {c\over 12}$. 

A bulk calculation where we only include smooth saddles thus can only reproduce the behaviour of CFTs if the one-point functions $\Gamma_i$ vanish for all operators with $dim(\phi_i)< {c\over 12}$. Although we are considering a bulk saddle-point calculation at large $c$, these one-point functions must vanish exactly to reproduce the bulk calculation, and not just at some order in $1/c$, or the exponential growth will eventually take over at sufficiently large $\beta$. This is a significant restriction on the CFT.  Indeed, as the one-point functions are determined by the consistency condition \eqref{const}, it is not clear if there exist CFTs which satisfy it.  In particular, we expect that solutions of \eqref{const} are uniquely determined up to a single
overall normalization. So it appears to be extraordinarily difficult to find a solution where the $\Gamma_i$ are non-vanishing only for heavy operators of dimension greater than $c/12$.

This restriction can be expressed in a natural way, by noting that the states with dimension less than ${c\over 12}$ are interpreted as perturbative states from the bulk point of view, while states with larger dimension are interpreted as black holes. In other words, in a theory with only smooth saddles, the cross cap state $|C\rangle $ is built purely out of black holes (states with dimension greater than $c\over 12$) rather than perturbative states. Theories dual to pure gravity are assumed to have no perturbative states other than the vacuum state. For such theories, what we have learned is that there is an additional restriction that the one-point function of the vacuum state (the $\RP^2$ partition function) vanish. 

\subsubsection{Minimal models}

Although our interest is mainly in theories at large central charge $c$, where the semi-classical bulk description is valid, the restriction on the one-point functions derived above motivates us to consider the minimal models, where the $\RP^2$ one-point functions can be computed explicitly.  
For example, in the Ising model we have \cite{Fioravanti:1993hf}
\begin{equation}
\Gamma_{\mathbb I}^2 = {\cal N} \frac{2 + \sqrt{2}}{2}, \quad \Gamma_{\epsilon}^2 = {\cal N} \frac{2 - \sqrt{2}}{2}, \quad \Gamma_{\sigma}^2 =0. 
\end{equation}
where ${\cal N}$ is an overall normalization.

The gravitational dual of the Ising model was considered in  \cite{Castro:2011zq}, where it was observed that the torus partition function of the Ising model (and other minimal models) can be reproduced by a sum over geometries.  It was argued that in the sum one had to include only the contributions from smooth metrics.  In this setup the quantization of the space of smooth metrics around thermal AdS gave the minimal model vacuum character, and the sum over handlebodies then gave the full Ising model partition function.

It is immediately clear that, once we consider non-orientable saddles, it is no longer possible to reproduce Ising model partition functions including only smooth geometries.  If we adopt the perspective of \cite{Castro:2011zq} -- namely that one should include only contributions from smooth metrics -- then the $\RP^2$ partition function immediately vanishes.  This is only possible if we set the normalization constant ${\cal N}=0$, which would imply the vanishing of all $\RP^2$ one point functions.  
In this case, according to (\ref{zis}) the Klein bottle partition function would vanish identically.  However, we have already identified a smooth saddle point which contributes to the Klein Bottle partition function.  

The calculation of the one-point functions can be extended to other minimal models, but we are not aware of explicit calculations in the literature. In the context of the pure gravity programme, it would be extraordinarily interesting to find a model with $\Gamma_i = 0$ for the operators with $dim(\phi_i)< {c\over 12}$, and some $\Gamma_i \neq 0$ for higher-dimension operators.  We conjecture, although we have not proven, that such theories do not exist.

\subsubsection{Including Singular Saddles}

In a general CFT, including the minimal models, the $\RP^2$ one point functions are generically non-zero, and we cannot reproduce the CFT behaviour by considering only smooth saddles in the bulk.  We are therefore led to consider the inclusion of singular saddle points.  
In the context of a pure gravity partition function, this inclusion of singular saddles would be a significant modification of the usual rules, and it is not clear what singularities should be considered acceptable. In the context of string theory, it might seem more natural to include such contributions, but note that the singularities are instanton contributions - they occur at points in the Euclidean time circle - so they do not correspond to including some additional ``particle-like'' degrees of freedom in the path integral as in \cite{Dijkgraaf:2000fq}. 

In section 2, we identified a class of natural singular bulk saddles for a non-orientable surface $\Sigma_g$,  where we have a smooth saddle for the orientable double cover $\hat \Sigma_g$, but the $\mathbb Z_2$ quotient has a fixed point in the bulk. For $\RP^2$, this gives the quotient $H^3/\Z_2$, which gives a contribution to $Z_{\RP^2}$.  We will obtain non-zero values for the one-point functions $\Gamma_i$ by including this saddle.   For the Klein bottle, we have in addition to the smooth saddle from the quotient of BTZ, a singular saddle from the quotient of thermal AdS, which we now include as a contribution to the Klein bottle partition function.  In the low temperature regime, it is this saddle that dominates the thermal partition function in the torus case. The action of the quotient will be half the action of the thermal AdS saddle, plus potentially a contribution from the singularities. Thus, 
\begin{equation}
I = I_{AdS}/2 + I_{sing} = - \beta \frac{c}{24} + I_{sing}. 
\end{equation}
Since the singularities are localised at points in the $t$ direction, their contribution should be independent of $\beta$ and will contribute an overall ($\beta$ independent) constant to the sum.  So this successfully reproduces the behaviour of the leading term in the CFT calculation \eqref{largelpf}. 

Thus, the inclusion of the singular $\mathbb Z_2$ quotients resolves the mismatch between general CFTs and the bulk results. In the next section, we will consider one-loop contributions to explore this match further and see if the bulk calculation can reproduce the Virasoro vacuum character appearing in \eqref{largelpf}.

\section{One-Loop Corrections}
\label{oneloop}

We now turn to the calculation of one-loop determinants around the bulk saddles, to further explore the agreement between bulk calculations and CFT expectations. Our main goal is to reproduce the Virasoro character in \eqref{zis} from the one-loop determinant around the saddle with fixed points, but we will also discuss the calculation of the one-loop determinant around the smooth saddle, where we find an interesting subtlety. We will first discuss the calculation using the method of images, following \cite{Giombi:2008vd}. We will then discuss the mode sum calculation, to clarify where the odd features in the smooth calculation come from. Our actual calculations are limited to a discussion of the one-loop determinant for a scalar field on the bulk saddle, but we comment on the expectations for the full gravity one-loop determinant. A discussion of other approaches to the one-loop determinant is relegated to the appendix. 

\subsection{Sum over images calculation}

We first consider the discussion using a sum over images calculation of the heat kernel, following the discussion in \cite{Giombi:2008vd}. We will consider the calculation for a scalar field in the bulk satisfying a massive wave equation, $\nabla^2 \phi - m^2 \phi =0$. The one-loop determinant on a bulk space $\mathcal M$ is obtained formally in terms of the eigenvalues $\lambda_n$ of the differential operator $\Delta = \nabla^2 - m^2$, 
\begin{equation}
\log \det \Delta = \sum_n \log \lambda_n. 
\end{equation}
We first approach this calculation by introducing the heat kernel, which is a sum over the eigenfunctions, 
\begin{equation}
K(t,x,y)= \sum_n e^{-\lambda_n t} \psi_n(x) \psi_n(y),
\end{equation}
where $\psi_n(x)$ is the eigenfunction of $\Delta$ with eigenvalue $\lambda_n$, and $x, y$ are points in $\mathcal M$. The one-loop determinant is then
\begin{equation}
\log \det \Delta = - \int_{0^+}^\infty \frac{dt}{t} \int_{\mathcal M} d^3x \sqrt{g} K(t,x,x). 
\end{equation}
The heat kernel is a useful tool because it can be defined as the a solution of the equation $(\partial_t + \Delta_x) K(t,x,y) = 0$ with the boundary condition $K(0,x,y) = \delta^3(x-y)$. In the present case, this is a particularly convenient way to approach the problem, as $\mathcal M = \mathbb H^3/\Gamma$, and we can obtain the heat kernel on $\mathcal M$ by method of images from the known solution on hyperbolic space, 
\begin{equation}
K^{\mathbb H_3/\Gamma} (t,x,y)= \sum_{\gamma \in \Gamma} K(t,x,\gamma y). 
\end{equation}
In \cite{Giombi:2008vd}, this was evaluated for thermal AdS. Here we will evaluate it for the bulk saddles for the Klein bottle introduced in section \ref{kleins} 
 
Let us consider first the quotient with fixed points, as this will give the simpler answer and it is the case we are most interested in testing. As discussed in section \ref{kleins}, this is a quotient of $\mathbb H^3$ by a discrete group $\Gamma \simeq \mathbb Z \times \mathbb Z_2$, where the $\mathbb Z_2$ generator is $\sigma$ and the $\mathbb Z$ generator is $\gamma$. We write the heat kernel on the quotient as the sum over images, 
\begin{eqnarray}
K^{\mathbb H^3/\Gamma}(t,x,x') &=& \sum_n K^{\mathbb H^3}(t,x,\gamma^n x') + \sum_n K^{\mathbb H^3}(t,x, \sigma \gamma^n x') \\ 
 &=&  K^{th}(t,x,x') + \sum_n K^{\mathbb H^3}(t,x, \sigma \gamma^n x'). \nonumber 
\end{eqnarray}
The first contibution is just the heat kernel on thermal AdS, as this is the quotient of $\mathbb H^3$ by the $\mathbb Z$ generated just by $\gamma$. Thus, the scalar  one-loop determinant is given by 
\begin{eqnarray}
- \log \det \Delta &=& \int_0^\infty \frac{dt}{t} \int_{\mathbb H^3/\Gamma} d^3x \sqrt{g} K^{\mathbb H^3/\Gamma}(t,x,x) \\ &=& - \frac{1}{2} \log \det \Delta_{th} + \int_0^\infty \frac{dt}{t} \int_{\mathbb H^3/\Gamma} d^3x \sqrt{g} \sum_{n} K^{\mathbb H^3}(t,x, \sigma \gamma^n x). \nonumber
\end{eqnarray}
The factor of half in the first term is from the smaller volume of the fundamental region in the quotient compared to thermal AdS. We need to calculate the second term explicitly. 
The heat kernel depends only on the geodesic distance 
\begin{equation}
r(x,x') = \mbox{arccosh}\left( 1 + \frac{ (y-y')^2 + |z-z'|^2}{2 yy'} \right). 
\end{equation}
Inserting 
\begin{equation}
y'  =e^{n \beta} \frac{y}{|z|^2 + y^2}, \quad z' =  - e^{n \beta} \frac{z}{|z|^2 + y^2}, 
\end{equation}
the geodesic length is 
\begin{eqnarray}
r(x,x') &=& \mbox{arccosh} \left( \frac{ y^2 (1 - e^{n \beta -2\rho})^2 + |z|^2 ( 1 + e^{n \beta - 2 \rho})^2 }{2 y^2 e^{n \beta - 2 \rho}} \right) \\ &=& \mbox{arccosh} \left( \cosh(n \beta + 2 \rho) + 2 \cot^2 \theta \cosh^2 (n \beta/2 + \rho) \right). \nonumber
\end{eqnarray}
As in \cite{Giombi:2008vd}, since the heat kernel is a function of $r$ it is convenient to trade the integral over $\theta$ in $d^3 x$ for an integral over $r$. The Jacobian  factor from the transformation will depend on $\rho$, so 
\begin{equation}
d^3 x \sqrt{g} = d\rho d\phi d\theta \frac{\cos \theta}{\sin^3 \theta} = \frac{d\rho}{4 \cosh^2(n \beta/2 + \rho)}  d\phi dr \sinh r. 
\end{equation}
The range of $r$ is $r \in (n \beta + 2 \rho, \infty)$. Recall the range of $\rho$ and $\phi$ in the fundamental region for the quotient is $\rho \in (0, \beta/2)$ and $\phi \in (0, 2\pi)$. The integral over $r$ is the one calculated in \cite{Giombi:2008vd}:
\begin{equation}
\int_{n \beta + 2 \rho}^\infty dr \sinh r K^{\mathbb H^3}(t,r) = \frac{ e^{-(m^2+1) t - \frac{(n \beta + 2 \rho)^2}{4 t}}}{4 \pi^{3/2} \sqrt{t}}. 
\end{equation}
In the determinant, the  integral over $\phi$ gives a factor of $2\pi$, but the integral over $\rho$ is non-trivial. Thus 
\begin{equation}
- \log \det \Delta =  - \frac{1}{2} \log \det \Delta_{th} + 2 \sum_{n} \frac{2\pi}{4} \int_0^{\beta/2} \frac{d\rho}{\cosh^2(n \beta/2 + \rho)} \int_0^\infty \frac{dt}{t}\frac{ e^{-(m^2+1) t - \frac{(n \beta + 2 \rho)^2}{4 t}}}{4 \pi^{3/2} \sqrt{t}}. 
\end{equation}
To do the $t$ integral we need to distinguish between $n \geq 0$ and $n <0$. The result is 
\begin{eqnarray}
- \log \det \Delta &=&  - \frac{1}{2} \log \det \Delta_{th} +  \sum_{n \geq 0} \frac{1}{8} \int_0^{\beta/2} \frac{d\rho}{\cosh^2(n \beta/2 + \rho)} \frac{e^{-\sqrt{1+m^2}(n \beta + 2 \rho)}}{n \beta/2 + \rho} \nonumber  \\ &&-  \sum_{n < 0} \frac{1}{8} \int_0^{\beta/2} \frac{d\rho}{\cosh^2(n \beta/2 + \rho)} \frac{e^{\sqrt{1+m^2}(n \beta + 2 \rho)}}{n \beta/2 + \rho}. 
\end{eqnarray}
Now a remarkable simplification occurs: we can replace the sum by a change of range in the $\rho$ integral. 
\begin{eqnarray}
- \log \det \Delta &=&  - \frac{1}{2} \log \det \Delta_{th} +  \frac{1}{8} \int_0^{\infty} \frac{d\rho}{\rho \cosh^2 \rho} e^{-\sqrt{1+m^2} \rho} -   \frac{1}{8} \int_{-\infty}^0 \frac{d\rho}{\rho \cosh^2 \rho} e^{\sqrt{1+m^2} \rho}  \nonumber  \\ &=&  - \frac{1}{2} \log \det \Delta_{th} +  \frac{1}{4} \int_0^{\infty} \frac{d\rho}{\rho \cosh^2 \rho} e^{-\sqrt{1+m^2} \rho}. 
\end{eqnarray}
The additional term is independent of $\beta$, so it is just some uninteresting constant normalisation factor in the one-loop determinant; the $\beta$ dependence is all in the piece that can be identified with the one-loop determinant on thermal AdS. 

The additional constant factor is divergent for the scalar. This is not unexpected; the quotient had fixed points, and one might expect a new UV divergence from propagators from the fixed point to itself in the sum over images. Indeed the contribution from $\rho =0$ in the two integrals correspond to the two fixed points at $y = 1, e^{\beta/2}$. 

We have not analysed the vector and metric one-loop determinants; these are much more difficult than in the thermal AdS calculation in \cite{Giombi:2008vd} because of non-trivial factors transforming the indices of the tensor fields when we act with $\sigma$. But we expect a similar logic will apply to them, so we can guess the form of the answer. The integrals over $r$ and $t$ are as in the thermal AdS case, but with the range for the $r$ integral changed as above, so in the output of these calculations $\beta$ will be replaced by $\beta + 2\rho$. There will then be a non-trivial $\rho$ integral, with the measure factor giving $\cosh(n \beta/2 + \rho)^{-2}$ in place of $|\sin(n \pi t)|^{-2}$. As a result, the answer for the full one-loop partition function should be
\begin{eqnarray}
\ln Z^{\mathrm{1-loop}}_{\mathrm{gravity}}  &=&  \frac{1}{2} Z^{\mathrm{1-loop}}_{th} + 2\pi \sum_n \int_0^{\beta/2} \frac{d \rho}{\cosh^2(n \beta/2 + \rho)}  \int_0^\infty \frac{dt}{t} \frac{e^{-\frac{(\beta+2\rho)^2}{4t}}}{4 \pi^{3/2} \sqrt{t}} [ e^{-t} - e^{-4t}] \nonumber \\ &=&  \frac{1}{2} Z^{\mathrm{1-loop}}_{th} - \frac{1}{2} \sum_{n \geq 0} \int_0^{\beta/2} \frac{d\rho}{(n\beta/2 + \rho) \cosh^2(n \beta/2 + \rho)} (e^{- 2(n \beta + 2 \rho)} - e^{-(n \beta + 2 \rho)}) \nonumber \\ && +\frac{1}{2}  \sum_{n< 0} \int_0^{\beta/2} \frac{d\rho}{(n\beta/2 + \rho) \cosh^2(n \beta/2 + \rho)} (e^{2(n \beta + 2 \rho)} - e^{(n \beta + 2 \rho)}) \nonumber \\ &=& \frac{1}{2} Z^{\mathrm{1-loop}}_{th} + \int_0^\infty \frac{d\rho}{\rho \cosh^2 \rho} e^{-2 \rho} (1- e^{-2\rho}). 
\end{eqnarray}
Again, the final contribution is independent of $\beta$, and gives a constant factor in the one-loop determinant. However, the prediction is that in the full calculation this factor would be finite. This may also not be surprising, as unlike the scalar theory, pure gravity has no physical bulk degrees of freedom to pick up new UV divergences at the fixed points. This gives some additional encouragement to think that we can sensibly deal with these singular saddles in pure gravity. 

If we believe our extrapolation from the scalar results, the one-loop partition function would then be
\begin{equation}
Z^{\mathrm{1-loop}}_{\mathrm{gravity}}  = C \sqrt{Z^{\mathrm{1-loop}}_{th}} = C \prod_{m=2}^\infty \frac{1}{(1 - e^{-m\beta})}. 
\end{equation}
This is precisely what we want; taken together with the fact that the leading saddle-point action will have half its thermal AdS value, this makes the partition function equal to the Virasoro character of the identity, 
\begin{equation}
Z_{\mathrm{gravity}}  = C e^{k \beta}  \prod_{m=2}^\infty \frac{1}{(1 - e^{-m\beta})}. 
\end{equation}
This is precisely the expected vacuum character, $\chi(i \beta/2\pi) = \chi(q = e^{-\beta})$. This is a strong test of the proposal that we need to include singular saddles in the bulk to reproduce the expected CFT behaviour. 

To see that the fact that we got a simple vacuum character in this calculation is quite non-trivial, it is useful to also consider the one-loop calculation for the smooth saddle. To do this by method of images, we write the smooth saddle as $\mathbb H^3 / \Gamma$ where $\Gamma$ is the $\mathbb Z$ group generated by $\kappa$ in \eqref{kappa}. The scalar heat kernel is
\begin{equation}
K^{\mathbb H^3/\Gamma}(t,x,x') = \sum_n K^{\mathbb H^3}(t,x,\kappa^n x').
\end{equation}
 Since $\kappa^2 = \gamma$, it is useful to break this sum into a sum over even and odd $n$:
\begin{eqnarray}
K^{\mathbb H^3/\Gamma}(t,x,x') &=& \sum_{n\ even} K^{\mathbb H^3}(t,x,\kappa^n x') +  \sum_{n\ odd} K^{\mathbb H^3}(t,x,\kappa^n x') \\ &=& \sum_{m} K^{\mathbb H^3}(t,x,\kappa^{2m} x') +  \sum_{n\ odd} K^{\mathbb H^3}(t,x,\kappa^n x')  \nonumber\\ &=& \sum_{m} K^{\mathbb H^3}(t,x,\gamma^{m} x') +  \sum_{n\ odd} K^{\mathbb H^3}(t,x,\kappa^n x') \nonumber \\ &=& K^{BTZ}(t,x,x') + \sum_{n\ odd} K^{\mathbb H^3}(t,x,\kappa^n x'), \nonumber
\end{eqnarray}
where we write the first term as the BTZ heat kernel as the $\gamma$ quotient in this case is interpreted as non-rotating BTZ. Thus the one-loop determinant is given by 
\begin{eqnarray}
- \log \det \Delta &=& \int_0^\infty \frac{dt}{t} \int_{\mathbb H^3/\Gamma} d^3x \sqrt{g} K^{\mathbb H^3/\Gamma}(t,x,x) \\ &=& - \frac{1}{2} \log \det \Delta_{BTZ} + \int_0^\infty \frac{dt}{t} \int_{\mathbb H^3/\Gamma} d^3x \sqrt{g} \sum_{n\ odd} K^{\mathbb H^3}(t,x,\kappa^n x). \nonumber
\end{eqnarray}
The factor of half in the first term comes again from the fact that the fundamental region has half the volume of the BTZ one. We need to compute the second term explicitly. For the geodesic distance
\begin{equation}
r(x,x') = \mbox{arccosh}\left( 1 + \frac{ (y-y')^2 + |z-z'|^2}{2 yy'} \right)
\end{equation}
for odd $n$, we take $y' = e^{n\beta/2} y$, $z' = e^{n \beta/2} \bar z$, which gives
\begin{equation}
r(x,x') = \mbox{arccosh}\left( \cosh \frac{n \beta}{2} + 2 \cot^2 \theta | \sin (\phi + i \frac{n\beta}{4}) |^2 \right). 
\end{equation}
As in \cite{Giombi:2008vd}, since the heat kernel is a function of $r$ it is convenient to trade the integral over $\theta$ in $d^3 x$ for an integral over $r$. The Jacobian factor from the change of variables is now $\phi$ dependent, so 
\begin{equation}
d^3 x \sqrt{g} = d\rho d\phi d\theta \frac{\cos \theta}{\sin^3 \theta} = d\rho \frac{d\phi}{ 4 | \sin (\phi + i \frac{n\beta}{4}) |^2} dr \sinh r. 
\end{equation}
The range of $r$ is $r \in (n \beta/2, \infty)$. The integral over $r$ was calculated in \cite{Giombi:2008vd}:
\begin{equation}
\int_{n \beta/2}^\infty dr \sinh r K^{\mathbb H^3}(t,r) = \frac{ e^{-(m^2+1) t - \frac{n^2 \beta^2}{16 t}}}{4 \pi^{3/2} \sqrt{t}}. 
\end{equation}
The integral over $\rho$ gives a factor of $\beta$, but the integral over $\phi$ is now non-trivial. Thus 
\begin{equation}
- \log \det \Delta =  - \frac{1}{2} \log \det \Delta_{BTZ} + 2 \sum_{n =1, 3, \ldots} \frac{\beta}{4} \int_0^\pi \frac{d\phi}{| \sin (\phi +   i \frac{n\beta}{4}) |^2} \int_0^\infty \frac{dt}{t}\frac{ e^{-(m^2+1) t - \frac{n^2 \beta^2}{16 t}}}{4 \pi^{3/2} \sqrt{t}}. 
\end{equation}
The integral over $t$ is also identical to the one evaluated in  \cite{Giombi:2008vd}, so 
\begin{equation}
- \log \det \Delta =  - \frac{1}{2} \log \det \Delta_{BTZ} +  \sum_{n =1, 3, \ldots} \frac{e^{-\frac{n}{2} \beta \sqrt{1+m^2}}}{2\pi n}  \int_0^\pi \frac{d\phi}{| \sin (\phi + i \frac{n\beta}{4}) |^2} .
\end{equation}
Up to this point the analysis is very similar to the singular case above, but in this case the second term really is a non-trivial function of $\beta$. The $\phi$ integral is easily evaluated to find 
\begin{equation}
- \log \det \Delta =  - \frac{1}{2} \log \det \Delta_{BTZ} +  \sum_{n =1, 3, \ldots} \frac{e^{-\frac{n}{2} \beta \sqrt{1+m^2}}}{2 n} \frac{1}{\sinh \frac{n\beta}{4} \cosh \frac{n\beta}{4}} .
\end{equation}
To make contact with the CFT side, it is convenient to define $q = e^{-\beta}$ as in  \cite{Giombi:2008vd}; then we have 
\begin{eqnarray}
- \log \det \Delta &=&  - \sum_{m=1}^\infty \frac{q^{2mh}}{m(1-q^m)^2}  +  2 \sum_{n =1, 3, \ldots} \frac{q^{nh}}{n (1-q^{n/2})(1 + q^{n/2})}  \\
&=&  - 2\sum_{n=2, 4, \ldots } \frac{q^{nh}}{n(1-q^{n/2})^2}  +  2 \sum_{n =1, 3, \ldots} \frac{q^{nh}}{n (1-q^{n/2})(1 + q^{n/2})} \nonumber \\ &=& 2 \sum_{n=1}^\infty \frac{q^{nh}}{n (1-(\sqrt q)^n)(1 - (- \sqrt q)^n)}, \nonumber
\end{eqnarray}
where $h = \frac{1}{2} (1 + \sqrt{1 + m^2})$. This is not just the square root of the BTZ answer, which we might have expected. 

We can extrapolate to guess the answer in the vector and metric cases, as we did for the singular saddle. The $r$ and $t$ integrals are identical to the ones in \cite{Giombi:2008vd}, and there is a non-trivial $\phi$ integral, which comes just from the measure factor in converting from $\theta$ to $r$, so it is independent of the spin of the field, and effectively multiplies the expression for each odd $n$ by $\tanh \frac{\beta}{4}$. Thus, the result for the one-loop gravity partition function is predicted to be 
\begin{eqnarray}
\ln Z^{\mathrm{1-loop}}_{\mathrm{gravity}}  &=&  \sum_{n=1}^\infty \frac{2 q^n (1- q^{n/2})}{n (1-(\sqrt q)^n)(1 - (- \sqrt q)^n)} \\ &=& 2 \sum_{n=1}^\infty \frac{2 q^n}{n (1 - (- \sqrt q)^n)}\nonumber \\ &=& - \sum_{m=2}^\infty \ln (1 - (- \sqrt q)^m)^2. \nonumber
\end{eqnarray}

\subsection{Mode sum calculation}

The result for the smooth saddle displayed some structure that is a little difficult to understand. To shed further light on this, we have calculated the one-loop determinant using a calculation from the  spectrum of the laplacian on the bulk saddle. We want to find the eigenfunctions $\Delta \psi_n = - \lambda_n \psi_n$, then we can write
\begin{equation}
K(t,x,y) = \sum_n e^{-\lambda_n t} \psi_n(x) \psi_n(y). 
\end{equation}

Here the idea is to choose a basis $\psi_n$ such that some elements are invariant under the quotient, so the heat kernel on the quotient space will be a sum over a subset of the modes considered in the original space. It is then obviously convenient to work with modes which in addition to being eigenfunctions of $\Delta$ are eigenfunctions of $\partial_t$, $\partial_\phi$ in the thermal AdS coordinates where the metric is 
\begin{equation}
ds^2 = \cosh^2 \rho dt^2 + d\rho^2 + \sinh^2 \rho d\phi^2, 
\end{equation}
with the periodic identification $t \sim t + \beta$. If we take
\begin{equation} \label{efn}
\psi = e^{im \frac{2\pi t}{\beta}} e^{in\phi} f_{mn}(\rho)
\end{equation}
for integers $m, n$, then the eigenvalue problem reduces to an ODE for $f$, 
\begin{equation}
(\cosh \rho \sinh \rho)^{-1} \partial_\rho ( \cosh \rho \sinh \rho \partial_\rho f) - \frac{4\pi^2}{\beta^2} m^2 \frac{f}{\cosh^2 \rho} - n^2 \frac{f}{\sinh^2 \rho} = - \lambda f.  
\end{equation}
We can reduce this to a hypergeometric equation by defining $z = \tanh^2 \rho$ and $\kappa = \frac{1}{2} - \frac{1}{2} \sqrt{1-\lambda}$, so $\lambda = - 4 \kappa (\kappa -1)$, and setting 
\begin{equation}
f = z^{\frac{n}{2}} (1-z)^\kappa F. 
\end{equation}
Then 
\begin{equation}
z(1-z) \partial_z^2 F + ((n+1) - (n+1+2\kappa) z) \partial_z F - (\frac{n^2}{4} + \frac{m^2 \pi^2}{\beta^2} + \kappa n + \kappa^2 ) F = 0. 
\end{equation}
the solution that is regular at $z=0$ is 
\begin{equation}
F = {}_2 F_{1}(\kappa + \frac{n}{2} + i m \frac{\pi}{\beta}, \kappa + \frac{n}{2} - im \frac{\pi}{\beta}; n+1 ;z). 
\end{equation}
The asymptotic expansion near the boundary at $z=1$ is given by the usual hypergeometric formula, 
\begin{eqnarray}
F(a,b;c;z) &=& \frac{\Gamma(c) \Gamma(c-a-b)}{\Gamma(c-a) \Gamma(c-b)} F(a,b;a+b-c+1;1-z) \\ &&+ \frac{\Gamma(c) \Gamma(a+b-c)}{\Gamma(a) \Gamma(b)} (1-z)^{1-2\kappa} F(c-a,c-b;c-a-b+1; 1-z).\nonumber
\end{eqnarray}
As a result, for all $\lambda <1$ the expansion has a non-normalizable component, so these values are not in the spectrum;  the Laplacian has no discrete part to its spectrum. For $\lambda >1$, by contrast $\kappa = \frac{1}{2} - i \alpha$, and the two asymptotic behaviours of $f$ are $(1-z)^{\frac{1}{2} \pm i \alpha}$, which are both acceptable, so any solution is normalizable. This is the continuous part of the spectrum of the Laplacian on hyperbolic space. Thus the spectrum of the Laplacian is all $\lambda > 1$. 

We want to choose a basis of solutions of the form \eqref{efn} which have nice transformation properties under the quotients we consider. for $n=m=0$, there's a single solution, and with the appropriate normalization, the eigenfunction is 
\begin{equation}
\frac{1}{2\pi \beta} f_{00\lambda}(\rho).
\end{equation}
For $n=0$, there are two solutions, 
\begin{equation}
\frac{1}{\pi \beta} \cos m \frac{2\pi}{\beta} t f_{m0\lambda}(\rho), \quad \frac{1}{\pi \beta} \sin m \frac{2\pi}{\beta}t  f_{m0\lambda}(\rho), \quad m >0 
\end{equation}
for $m=0$, there are two solutions,  
\begin{equation}
\frac{1}{\pi \beta} \cos n \phi f_{0n\lambda}(\rho), \quad \frac{1}{\pi \beta} \sin n\phi f_{0n\lambda}(\rho), \quad n >0,
\end{equation}
and for general $m,n$ there are four solutions, 
\begin{equation}
\frac{2}{\pi \beta} \cos m \frac{2\pi}{\beta} t \cos n \phi f_{mn\lambda}(\rho), \frac{2}{\pi \beta} \cos m \frac{2\pi}{\beta} t \sin n \phi f_{mn\lambda}(\rho)\quad m,n >0,
\end{equation}
\begin{equation}
\frac{2}{\pi \beta} \sin m \frac{2\pi}{\beta} t \cos n \phi f_{mn\lambda}(\rho), \frac{2}{\pi \beta} \sin m \frac{2\pi}{\beta} t \sin n \phi f_{mn\lambda}(\rho) \quad m,n >0,
\end{equation}

Plugging this into the general formula, we can write the scalar heat kernel on thermal AdS for coincident points  as 
\begin{equation}
K(t,x,x) = \frac{1}{2 \pi \beta} \int_1^\infty d\lambda \mu(\lambda) e^{-\lambda t} ( f_{00\lambda}^2 + 2 \sum_{m>0} f_{m0\lambda}^2 + 2 \sum_{n >0} f_{0n\lambda}^2 + 4 \sum_{m,n >0} f_{mn\lambda}^2 ), 
\end{equation}
where $\mu(\lambda)$ is the spectral function, which measures the degeneracy of modes at each value of $\lambda$. This can be determined by applying the normalization condition 
\begin{equation} \label{muint}
\int d \rho \cosh \rho \sinh \rho f_{mn\lambda} (\rho) f_{m'n'\lambda'} (\rho) = \delta_{mm'} \delta_{nn'} \frac{\delta(\lambda - \lambda')}{\mu(\lambda)}. 
\end{equation}
This normalization condition can be applied for any choice of $m, n$; in particular, it can be applied for $m =m'=0$, which implies that the spectral function $\mu(\lambda)$ is independent of the temperature. As a result, to understand the temperature dependence of the heat kernel, we will not need to know the spectral function explictly. A discussion of this spectral function can be found for example in \cite{Camporesi:1994ga}.

Note that the heat kernel on thermal AdS is manifestly independent of $t, \phi$, which is a consequence of the translation invariance in these directions, but is some non-trivial function of $\rho$. It is the dependence on $\rho$ which should make these sums convergent. The 1,2,2,4 structure in the sum may seem a little strange; this looks more natural if we observe that what it means is that the heat kernel is naturally written as the sum 
\begin{equation} \label{BTZk}
K(t,x,x) = \frac{1}{2 \pi \beta} \int_1^\infty d\lambda \mu(\lambda) e^{-\lambda t} \sum_{m,n \in \mathbb Z} f_{m n \lambda}^2 . 
\end{equation}

Now consider the quotient under $t \to - t, \phi \to \phi + \pi$, to obtain the expression on the singular quotient. For even $n$ (including $n=0$), the $\cos t$ modes are invariant, and for odd $n$, the $\sin t$ modes are invariant. For $m=0$, the modes with even $n$ are invariant. The $n=m=0$ mode is also invariant. On the quotient, the volume of the $\phi, t$ space is halved, so the normalization factors in the modes in the quotient is different by a factor of $\sqrt{2}$. Thus the heat kernel on the quotient is 
\begin{eqnarray} \nonumber 
K^{sing}(t,x,x) &=& \frac{1}{\pi \beta} \int_1^\infty d\lambda \mu(\lambda) e^{-\lambda t} (f_{00\lambda}^2 +2 \sum_{m>0} f_{m0\lambda}^2 \cos^2 m \frac{2\pi}{\beta} t + 2 \sum_{n >0, even} f_{0n\lambda}^2 \\ &&+ 4 \sum_{m,n >0, n\ even} f_{mn\lambda}^2 \cos^2 m \frac{2\pi}{\beta} t + 4 \sum_{m,n >0, n\ odd} f_{mn\lambda}^2 \sin^2 m \frac{2\pi}{\beta} t ).
\end{eqnarray}
This is now a function of $t$, which is not surprising, as the quotient broke homogeneity in the $t$ direction. The spectral function $\mu(\lambda)$ here is unchanged, as it is determined by the same normalization integral \eqref{muint}. 

Integrating over $t$ and $\phi$, 
\begin{equation}
\int dt d\phi  K^{sing} (t,x,x) = \int_1^\infty d\lambda \mu(\lambda) e^{-\lambda t} (f_{00\lambda}^2+  \sum_{m>0} f_{m0\lambda}^2  + 2 \sum_{n >0, even} f_{0n\lambda}^2 + 2 \sum_{m,n >0} f_{mn\lambda}^2). 
\end{equation}
To understand the sum over images result, we compare this to the result of integrating the thermal heat kernel over the fundamental region; that is, we define $\Delta K = K^{sing}(t,x,x) - K(t,x,x)$ where $K(t,x,x)$ is given by \eqref{BTZk}. Then 
\begin{equation}
\int dt d\phi  \Delta K(t,x,x) = \int_1^\infty d\lambda \mu(\lambda) e^{-\lambda t} (\frac{1}{2} f_{00\lambda}^2 + \sum_{n >0, even} f_{0n\lambda}^2  -\sum_{n >0, odd} f_{0n\lambda}^2). 
\end{equation}
Because the difference only involves terms with $m=0$, it is manifestly independent of the modular parameter. This explains the miraculous-seeming combination of integrals in the sum over images calculation which gave a constant result.

The smooth quotient which gives the geon is $t \to t + \beta/2$, $\phi \to -\phi$ in these coordinates. Thus, the modes which are invariant are $\cos n \phi$ for even $m$ (including $m=0$), and $\sin n \phi$ for odd $m$. For $n=0$ even $m$ is invariant.  The $n=m=0$ mode is also invariant. Thus the heat kernel on the quotient is 
\begin{eqnarray} \nonumber 
K^{smooth}(t,x,x) &=& \frac{1}{\pi \beta} \int_1^\infty d\lambda e^{-\lambda t} (f_{00\lambda}^2 +2 \sum_{m>0, even} f_{m0\lambda}^2 + 2 \sum_{n >0} f_{0n\lambda}^2  \cos^2 n \phi \\ &&+ 4 \sum_{m,n >0, m\ even} f_{mn\lambda}^2 \cos^2 n \phi + 4 \sum_{m,n >0, m\ odd} f_{mn\lambda}^2 \sin^2 n \phi ).
\end{eqnarray}
Integrating over the fundamental region,  
\begin{equation}
\int dt d\phi  K^{smooth}(t,x,x) = \int_1^\infty d\lambda \mu(\lambda) e^{-\lambda t} (f_{00\lambda}^2+ 2 \sum_{m>0,even} f_{m0\lambda}^2  +  \sum_{n >0} f_{0n\lambda}^2 + 2 \sum_{m,n >0} f_{mn\lambda}^2). 
\end{equation}
Defining similarly $\Delta K(t,x,x) = K^{smooth}(t,x,x) - K(t,x,x)$, where $K(t,x,x)$ given by \eqref{BTZk} is now interpreted as the heat kernel on BTZ, 
\begin{equation}
\int dt d\phi \Delta K(t,x,x) = \int_1^\infty d\lambda \mu(\lambda) e^{-\lambda t} (\frac{1}{2} f_{00\lambda}^2+\sum_{m>0,even} f_{m0\lambda}^2 -\sum_{m>0,odd} f_{m0\lambda}^2). 
\end{equation}
This can also be rewritten, perhaps more suggestively, as 
\begin{equation}
\int dt d\phi \Delta K(t,x,x) = \int_1^\infty d\lambda \mu(\lambda) e^{-\lambda t} \frac{1}{2} \sum_{m \in \mathbb Z} (-1)^m f_{m0\lambda}^2. 
\end{equation}
So we see that there is a non-trivial difference coming from zero mode contributions on the spatial circle. In this case, this leads to a non-trivial temperature dependence in the additional contribution to the one-loop determinant. The key difference between the singular and smooth saddles is that in the former case the difference involves zero modes in the time direction, which are independent of the modulus, but in the latter case it involves zero modes in the spatial direction, which depend on the modulus. 

Here again we have only carried out the calculation for scalar fields, but it should be more straightforward to extend this analysis to the vector and metric to verify that the contributions on the singular saddle are the same as in thermal AdS, up to an additional contribution independent of the modulus.

\subsection{One-loop corrections: CFT calculation}
\label{oneloopd} 

The singular saddle reproduced the expected behaviour of the CFT partition function at low temperatures (large modular parameter $\beta$), as the one-loop determinant around the quotient was just the square root of the one-loop determinant around the thermal AdS solution. This is a useful confirmation of our argument that we need to include this singular saddle to reproduce the expected CFT behaviour. However, the one-loop determinant around the smooth saddle did not have such a simple form. We might naively have expected this just to be given by the Virasoro character in the other channel, as it is in the torus case. Here we examine this issue more carefully from the CFT point of view and see that in fact we would expect a non-trivial contribution from zero modes in the CFT analysis, just as we are finding in the bulk calculation. 

The CFT partition function at small modular parameter $\beta$, where we would expect the smooth saddle to dominate, is conveniently expressed as in \eqref{Ppf} as a trace over states with an insertion of a parity projection. The bulk calculation we have considered is for a scalar field, which corresponds to the contribution of the dual scalar operator $\phi$ in the CFT partition function. 

It is useful to first recall the torus partition function. The Fock space basis of states for the scalar is labelled by a string of non-negative integers $n_{\ell \ell'}$ for $\ell, \ell' = 0, \ldots \infty$, so the basis states are 
\begin{equation}
|\psi_{\{ n_{\ell \ell'}\}} \rangle = \prod_{\ell, \ell' = 0}^\infty (\partial^\ell \bar \partial^{\ell'} \phi)^{n_{\ell \ell'}} | 0 \rangle. 
\end{equation}
The partition function on the rectangular torus is given by a sum over this basis, 
\begin{equation}
\mbox{Tr} (e^{-\beta H}) = \prod_{\ell, \ell' = 0}^\infty \sum_{n_{\ell \ell'} = 0}^\infty e^{-\beta (2h + \ell + \ell') n_{\ell \ell'}} =  \prod_{\ell, \ell' = 0}^\infty \frac{1}{1- e^{-\beta (2h+\ell + \ell')}}. 
\end{equation}
For comparison to the Klein bottle partition function, this can be written as 
\begin{equation} \label{tpn}
\mbox{Tr} (e^{-\beta H}) =  \prod_{\ell =0}^\infty \frac{1}{1- e^{-\beta (2h+2\ell)}} \prod_{\ell = 1}^\infty \prod_{\ell' = 0}^{\ell-1} \frac{1}{(1 - e^{-\beta (2h + \ell +  \ell')})^2} ,
\end{equation}
where we have separated out the diagonal and off-diagonal contributions, and made use of the fact that the energy only depends on $\ell + \ell'$ to combine contributions above and below the diagonal. 

For the Klein bottle, the partition function includes an action of parity. This means we should organise our basis into parity eigenstates. In the basis above, the states where $\{n_{\ell \ell'} \}$ is invariant under interchange of $\ell$ with $\ell'$ are parity-even, and the others are not parity eigenstates. From the non-invariant basis states we can construct a parity-even combination 
\begin{equation}
|\psi_+ \rangle = |\psi_{\{ n_{\ell \ell'}\}} \rangle +  |\psi_{\{ n_{\ell' \ell}\}} \rangle
\end{equation}
and a parity-odd combination 
\begin{equation}
|\psi_- \rangle = |\psi_{\{ n_{\ell \ell'}\}} \rangle -  |\psi_{\{ n_{\ell' \ell}\}} \rangle.
\end{equation}
Since these states have the same energy, their contribution to the trace will cancel, and we are left with just the contribution from the invariant states where $\{n_{\ell \ell'} \} = \{n_{\ell' \ell} \}$, that is, those labelled by strings $\{n_{\ell \ell'} \}$ with $n_{\ell_1 \ell_2} = n_{\ell_2 \ell_1}$. 

That is, for the partition function \eqref{Ppf} the non-zero contribution comes from the trace over states where we act with the same number of $\partial^\ell \bar \partial^{\ell'} \phi$ and $\partial^{\ell'} \bar \partial^\ell \phi$, to form basis states
\begin{equation}
\prod_{\ell =0  }^\infty (\partial^\ell \bar \partial^\ell \phi)^{n_{\ell}}  \prod_{\ell =1}^{\infty} \prod_{\ell'=0}^{\ell-1} (\partial^\ell \bar \partial^{\ell'} \phi  \partial^{\ell'} \bar \partial^\ell \phi)^{n_{\ell \ell'}} | 0 \rangle. 
\end{equation}
We see already that the diagonal terms with $\ell = \ell'$, corresponding to the momentum zero modes, have a different behaviour to the off-diagonal terms. 

Evaluating the partition function, 
\begin{equation}
\mbox{Tr}(P e^{-\frac{\beta}{2} H}) = \prod_{\ell =0  }^\infty \frac{1}{1 - e^{-\frac{\beta}{2}(2h + 2\ell)}} \prod_{\ell = 1}^\infty \prod_{\ell' = 0}^{\ell-1} \frac{1}{1 - e^{-\beta (2h + \ell +  \ell')}} . 
\end{equation}
This has the same qualitative structure seen in the gravity calculation; the off-diagonal part is the square root of the torus partition function \eqref{tpn}, but the diagonal part, corresponding to the momentum zero modes, spoil this pattern.

\section*{Acknowledgements}

We are grateful for discussions with Cindy Keeler, Matthias Gaberdiel, Henry Maxfield and Shamit Kachru.  AM is supported by the National Science and Engineering Council of Canada and by the Simons Foundation. SFR is supported by STFC under grant number ST/L000407/1. 
\appendix

\section{Alternative approaches to the one-loop determinant}

In our analysis of the one-loop determinant, we only analysed the scalar modes explicitly, as the extension to vector and metric was non-trivial. Here we would like to comment on two other approaches to calculating the one-loop determinant used in the literature where one might have hoped that the extension to vector and metric would be more straightforward, and explain why we were not able to use them. 

\subsection{More algebraic approach}

The work of \cite{Giombi:2008vd} was generalised by \cite{David:2009xg}, who make more use of group theory and the description of the hyperbolic plane as a group manifold. A central element in their discussion is the construction of a section of the principal bundle over the hyperbolic plane which is invariant under the quotient. This relies on the transformation we quotient by being expressible as 
\begin{equation}
g \to A g B^{-1} 
\end{equation}
for some group elements $A$, $B$. (See their (2.24), and (4.2) for the explicit representation). Unfortunately, the quotients which give non-orientable boundaries don't seem to be expressible in this form. The obstruction is easiest to see if we consider the quotient of $\mathbb H^3$ to obtain a space with $\mathbb R P^2$ boundary: in terms of the embedding coordinates $T, X_i$, the global coordinates on the hyperbolic space are
\begin{equation}
T = \cosh \chi, X_i = \sinh \chi x_i,
\end{equation}
where $x_i$ are coordinates on a unit $S^2$. So the quotient that turns the boundary $S^2$ into $\mathbb RP^2$ is reversing the sign of the $x_i$. In terms of the embedding coordinates, the $SL(2, \mathbb C)$ group element is
\begin{equation}
g = \left( \begin{array}{cc} T+X_1 & X_2 + i X_3 \\ X_2 - i X_3 & T - X_1 \end{array} \right).
\end{equation}
The quotient is thus not $ g \to -g$ as in the $S^3$ case, but sends
\begin{equation}
g = \left( \begin{array}{cc} \alpha & \gamma \\ \kappa & \delta \end{array} \right)
\end{equation}
to
\begin{equation}
g' = \left( \begin{array}{cc} \delta & - \gamma \\ -\kappa & \alpha \end{array} \right).
\end{equation}
For this to be of the form $g' = A g B^{-1}$ for some $A, B$ requires
\begin{equation}
\delta = B_{11} (A_{11} \alpha + A_{12} \kappa) + B_{21} (A_{11} \gamma + A_{12} \delta), 
\end{equation}
\begin{equation}
-\gamma = B_{12} (A_{11} \alpha + A_{12} \kappa) + B_{22} (A_{11} \gamma + A_{12} \delta), 
\end{equation}
\begin{equation}
-\kappa = B_{11} (A_{21} \alpha + A_{22} \kappa) + B_{21} (A_{21} \gamma + A_{22} \delta), 
\end{equation}
\begin{equation}
\alpha = B_{12} (A_{21} \alpha + A_{22} \kappa) + B_{22} (A_{21} \gamma + A_{22} \delta). 
\end{equation}
Now for this to hold for any $\alpha, \gamma, \kappa, \delta$, the first equation requires e.g. $B_{11} A_{11} = 0$, but the second requires $B_{22} A_{11} = -1$, and the third requires $B_{11} A_{22} = -1$, which produces a contradiction.

\subsection{Quasi-normal modes}

Another approach would be to follow \cite{Denef:2009kn}, and evaluate the one-loop determinant from the quasi normal mode frequencies. For the quotient with fixed points, it's useful to think of the space as thermal AdS, so the mode spectrum is 
\begin{equation}
z_{n,l,\pm} = \pm (2n + l + \Delta), \quad n= 0,1,2, \ldots ,\quad  l \in \mathbb Z, 
\end{equation}
and their expression for the one-loop determinant is 
\begin{equation}
Z = e^{\textrm{Pol}(\Delta)} \prod_{z_\star} \frac{1}{2 \sinh \frac{|z_\star|}{2T}} = e^{\textrm{Pol}(\Delta)} \prod_{n,l} \frac{1}{4 \sinh^2 \frac{|z_{n,l}|}{2T}} 
\end{equation}
In taking the quotient under $t \to - t$, $\phi \to \phi + \pi$ we should keep the normal modes in this spectrum which are invariant under the quotient. For even $l$, these are the symmetric combination of the modes labelled by $z_{n,l,\pm}$ and for odd $l$ it's the antisymmetric combination of the modes labelled by $z_{n,l,\pm}$. Thus for each $n, l$, we keep one of the two modes. Thus the one-loop determinant on the quotient is 
\begin{equation}
Z = e^{\textrm{Pol}(\Delta)} \prod_{n,l} \frac{1}{2 \sinh \frac{|z_{n,l}|}{2T}} 
\end{equation}
So up to possible differences in the polynomial part, the one-loop determinant on the quotient is precisely the square root of that on thermal AdS. 

For the geon quotient, this basis of normal modes is not useful, as none of them are invariant under $t \to t + \beta/2$, $\phi \to -\phi$. Instead we should use the basis of quasinormal modes on BTZ, where the spectrum is 
\begin{equation}
z_{p,s} = p - 2\pi Ti (\Delta + s), \quad s = 0,1,2, \ldots ,\quad  p \in \mathbb Z, 
\end{equation}
and $\bar{z}_{p,s} = z_{p,s}^*$. Now to get an invariant combination we take a combination of $z_{p,s}$ and $\bar{z}_{-p,s} = -z_{p,s}$: again the symmetric combination for $p$ even and the antisymmetric combination for $p$ odd. 

The one-loop determinant on BTZ was
\begin{equation}
Z = e^{\textrm{Pol}(\Delta)} \prod_{z_\star} \frac{\sqrt{z_\star \bar z_\star}}{4\pi^2 T} \Gamma( \frac{i z_\star}{2\pi T}) \Gamma( \frac{-i \bar z_\star}{2\pi T}) =  e^{\textrm{Pol}(\Delta)} \prod_{p,s} \frac{i z_{p,s}}{4\pi^2 T} \Gamma( \frac{i z_{p,s}}{2\pi T})^2,
\end{equation}
and up to changes in the polynomial part, the one-loop determinant on the geon would be the square root of this, 
\begin{equation}
Z =   e^{\textrm{Pol}(\Delta)} \prod_{p,s} \frac{i \sqrt{z_{p,s}}}{4\pi^2 T} \Gamma( \frac{i z_{p,s}}{2\pi T}).
\end{equation}
However, we know that there {\it is} a non-trivial difference between the one-loop determinant on the quotient and the square root of BTZ. In this approach to the calculation, this difference is hidden in the Pol$(\Delta)$ prefactor, so we do not get much control of it. This calculation thus provides a nice illustration of the subtleties in applying the quasi-normal mode approach, and while it would in principle be straightforward to extend this approach to vector and metric fields, it would be better to do so in the full mode sum analysis where the zero mode contributions can also be controlled.

\bibliographystyle{JHEP}
\bibliography{klein}

\end{document}